%% file: art.tex
\documentclass[12pt]{article}
\usepackage{geometry}
\usepackage{fancyhdr}
\usepackage{multicol}
\usepackage[perpage,stable,symbol]{footmisc}
\usepackage{graphics}
\usepackage{slatex}
\input{slinit}
\begin{document}

\title{Using a hierarchy of Domain Specific Languages in complex
software systems design}
\author{V.~S.~Lugovsky {\tt <VSLougovski@lbl.gov>}}

\maketitle

\begin{abstract}
A new design methodology is introduced, with some examples on building Domain Specific Languages hierarchy on top of Scheme.
\end{abstract}

\begin{multicols}{2}


\section{Introduction}

Programs that write programs that write programs (...) Too complicated? A hackers technique which can not be applied to the ``real world'' 
problems? This is exactly how IT industry specialists think about metaprogramming. And this is a completely wrong notion! 

Metaprogramming is the {\it only} known way to reduce
the complexity significantly. In some areas ``programs that write programs''
are accepted by the industry due to the enormous level of complexity
of the corresponding handwritten code --- regular expressions, 
lexers and 
parsers generators to name a few, and code wizards and templates in
popular ``integrated development environments'' are also widely used. 
But this does not help at all in the
overall methodology recognition. The industry's most beloved and widely
buzzworded language, Java,  does not have even  such a rudimentary 
preprocessor as C does. Very few C++ programmers
have an idea on how to use the templates, they just utilize STL without
any understanding of the true source of the power. Even in the enlightened
world of Lisp programming the misunderstanding is surprisingly wide: almost
all of Lisp dialects and Scheme implementations have problems with macros
(not so many people are using them), and even the current 
Scheme standard R5RS contains only hygienic macros that can hardly be 
recognized as ``true'' macros as they hide an access to the host language.

 This situation looks like a paradox. On the one hand, industry uses the
metaprogramming ideas and tools, and it is easy to imagine how it would 
suffer without
them. On the other hand, industry does not want to hear anything related
to the metaprogramming. It does not want people inventing new programming 
languages --- plenty of industry coders barely use only one language
and IT managers believe without any reason that they can not be taught 
to use more \cite{python}. 

Industry prefers to ``re--invent a wheel'' and to express any sort of  complexity
in the form of libraries for static, steady languages. For some strange reason learning complicated libraries for a language  which
barely fits problem domain needs is preferred to learning a {\it small} new language specifically designed for it.

In this paper I am trying to advocate the 
metaprogramming approach as the major design methodology for complex 
systems. Sounds like another one ``silver bullet'' invention? There were
many methodologies claiming to solve all possible problems of the mankind ---
RUP, eXtreme programming, etc. Why do we need another one? Simply because
the previous approaches did not succeed. They were too tied to 
particular programming technologies (mostly --- OOP varieties), which are
definitely not ``silver bullets''. Metaprogramming methodology
is different, it strongly encourages the use of all possible 
programming technologies and to invent the ``impossible'' ones.

\section{Domain specific languages}

 Below I am providing an outline of the proposed methodology. 

 Any problem domain can be best expressed using a language (mathematical,
programming, natural, ...) specially designed for it. In most cases there
should be one entity in a language for every entity in a problem domain.
For example, if a problem domain is the recognition of syntax constructions in
a characters stream, the Domain Specific Language should contain
characters and characters sets as a primary entity and automata constructions
for expressing syntax. That is enough --- regular expressions language is designed. It 
is hard to believe that somebody
will ever invent anything better for this purpose than 
this ``most optimal'' DSL.

 If a problem domain is already 
specified as an algebra, we even do not have to design the DSL: it will be
this algebra itself, galvanised with any underlying computational semantics ---
this is the way SQL was born. If a problem domain is 3D graphics, linear algebra and stereometry should be used. All the
languages and data formats dedicated to 3D contain subsets of this
formal theories. 

As it is stated in \cite{dsls},
\begin{quote}
``The object of a DSL-based software architecture is to minimise 
the semantic distance between the system's specification and its 
implementation.'' 
\end{quote}

\section{Core language}

 For any problem it is convenient to have a language that best fits it. There already exist specialized languages  for
some common problems. But what
to do if none is available? The answer is trivial: implement it. Implementation
of domain specific languages is not very tricky. 
An approach I will describe here is based on metaprogramming
techniques. It requires a so called {\it Core Language}, on top
of which we will build a hierarchy of our domain specific languages. The Core
Language should possess the following properties:
\begin{itemize}
\item True macros. That is, we must have an access to a complete 
programming language (preferably the same as a host language, or
a different one) inside the macro definitions. Macros should be real programs
which can do anything that the programs written in the host language can do. 
Macros are producing the code in the host language, in the form of text or directly
as an abstract syntax tree.
\item True runtime eval. Programs that are generated in the runtime should
 be evaluated. This can be a different language than the host language, or,
 better, the same one.
\item Turing-completeness. This should be a real programming language, 
  equivalent in its expressive power to the ``general purpose'' languages.
\item Simplicity. It is an extensible core and should not contain any 
unnecessary complexity that can be later added by a user who {\it 
really} needs it.
\item Comprehensive and easy to use data types system. If a type system
is well suited for expressing any possible abstract syntax trees, the language
fits this requirement.
\end{itemize}

 On top of the Core Language we have to build  functionality that will
be needed to implement programming languages. It is lexing, parsing,
intermediate languages that fit well computational models different
from the model of the Core Language (e.g., if the core language is imperative 
or an eager functional, we will need a graph reduction engine to implement
lazy functional DSLs, or a term unification engine to implement logical
languages and a stack machine if we have to go to lower levels). The
Core Language enriched with this  ``Swiss army knife'' for programming languages 
development then becomes a major tool for any
project.

\section{New methodology}

The development process must fit in the following chain:
\begin{itemize}
\item divide the problem into sub--problems, possibly using some object
oriented design techniques, or whatever fits better.
\item formalize each sub--problem.
\item implement the Domain Specific Language after this formalization, using
the Core Language and other DSL with the same semantics.
\item solve the problem using the best possible language.
\end{itemize}
This way any project will grow into a tree (hierarchy) of domain specific
languages. Any language is a subset or a superset of another
language in the hierarchy (or, may be, combination of several
languages), and the amount of coding for a new language if we already have
a deep and comprehensive hierarchy is quite small.

 A development team working within this methodology should consist
 of at least one specialist who maintains this hierarchy, an architect
 who formalizes problems, and a number of coders who
 specialize in particular problem domains, they even may not be
  programmers at all --- they just have to know well their domains
 and operate them in terms that are as close as possible to the native
 problem domain terminology. For example, HTML designer will be happy
 operating HTML--like tags for his templates (that is why JSP
 custom tags are so popular); mathematician will find  a
 language modelled after the standard mathematical notation intuitive --- for this reason
 Wolfram Mathematica is so popular among non-programmers; game script
 writer will ope\-rate a language expressing characters, their
 properties and action rules --- {\it stating}, not {\it
 programming}. This list can be continued infinitely.

\section{Scheme example}

 A good example of a practical Core Language is
 Scheme (with addition of Common Lisp--style macros). 
 It uses S--expressions as an AST, and
 S--expressions composition is very natural. S--expressions are good
 enough to represent any possible AST (for example, XML is naturally 
 represented as SXML). It provides a true runtime eval hosting the
 same language as in compile time. There exist some practical and
 efficient Scheme implementations which provide performance
 acceptable for most tasks, good FFI, and, thus, integration with
 legacy libraries.

 Let us start with adding the functionality described
 above to Scheme. First of all we will need parsing --- not all of our team
 members are fond of parentheses, so we have to implement many
 complicated syntaxes. The most natural way for a functional
 programming language is to implement a set of parsing combinators for
 building recursive descendant parsers (mostly LL(1), but it is
 not such a fixed limit as LALR(1) for Yacc--like automata
 generators).

 Of course we will use metaprogramming wherever possible. All the parsers
 should be functions which consume a list of tokens
 (e.g. characters) as an input and return the result in the following
 form:
\begin{schemedisplay}
((RESULT anyresult) unparsed-input-rest)
 or
((FAIL reason) input)
\end{schemedisplay}

 To access the parsing result we will provide the following macros:
\begin{schemedisplay}
(define-macro (success? r)
  `(not (eq? (caar ,r) 'FAIL)))
\end{schemedisplay}

 And if we are sure that we have some result, we will use the following
 macro to extract it (otherwise, this will return a fail message):
\begin{schemedisplay}
(define-macro (result r)
  `(cdar ,r))
\end{schemedisplay}

 In any case, we can access the rest of the stream after the parsing pass:
\begin{schemedisplay}
(define-macro (rest r)
  `(cdr ,r))
\end{schemedisplay}

 These macros could also be implemented as functions. But
all the macros are available in the context of macro definitions while
functions are not.

 Almost all of the parsers should fail on the end of the input, so the
 following safeguard macro will be extremely useful:
\begin{schemedisplay}
(define-macro (parser p)
  `(lambda (l)
    (if (null? l) '((FAIL "EMPTY"))
         (,p l))))
\end{schemedisplay}

 Now this game becomes more interesting. Here is a very handy macro that
 nests a sequence of applications 
 into the form of \scheme|(m p1 (m p2 ... (m px pn)))|:
\begin{schemedisplay}
(define-macro (pselect m p1 . po)
 (if (null? po) p1
     (let ((p2 (car po))
           (px (cdr po)))
      `(,m ,p1 (pselect ,m ,p2 ,@px)))))
\end{schemedisplay}

 Sequence parsing combinator with two arguments can be declared as
 follows:
\begin{schemedisplay}
(define-macro (p+0 p1 p2)
 `(lambda (l)
   (let ((r1 (,p1 l)))
    (if (success? r1)
     (let ((r2 (,p2 (rest r1))))
      (if (success? r2)
          (cons (cons 'RESULT 
                  (append 
                    (result r1) 
                    (result r2)))
                (rest r2))
          (cons (list 'FAIL "p+" (car r2)) l)))
     r1))))
\end{schemedisplay}

 And it will be immediately turned into the sequence parsing
 combinator with an arbitrary number of arguments:
\begin{schemedisplay}
(define-macro (p+ p1 . po)
  `(pselect p+0 ,p1 ,@po))
\end{schemedisplay}

 The last definition looks surprisingly compact, thanks to the
 \scheme|pselect| macro. From this stage the power of metaprogramming
 becomes more and more obvious.

 Just as a reference, we will show here the definition of a choice 
combinator:
\begin{schemedisplay}
(define-macro (pOR0 p1 p2)
 `(lambda (l)
   (let ((r1 (,p1 l)))
         (if (success? r1)
              r1
              (,p2 l)))))
\end{schemedisplay}

 And its nested version is obvious:
\begin{schemedisplay}
(define-macro (pOR p1 . po)
 `(pselect pOR0 ,p1 ,@po))
\end{schemedisplay}

 We will skip the rest of the combinators definitions and just show
 what we gained after all. For example, now to define a floating
 point number recognizer, we can use this definition:
\begin{schemedisplay}
(define parse-num
  (p+
    (pOR (pcsx-or (#\- #\+))
         parse-any)
    (pMANY pdigit)
    (pOR (p+ (pcharx #\.)
             (pMANY pdigit))
         parse-any)))
\end{schemedisplay}

 It looks like BNF, but still too Schemish. This is already a Domain
 Specific Language on top of Scheme, but it does not conform to the 
 perfectionist requirement. However, we can use this still not perfect
 parsing engine to implement an intermediate regular expressions
 language as a macro. Omitting the definitions, we will show the
 previous recognizer implemented in a new way:

\begin{schemedisplay}
(define parse-num
   (regexp
     ((#\- / #\+) / parse-any) + 
      (pdigit *) +
      (("." + (pdigit *)) / 
      parse-any)))
\end{schemedisplay}

 This new Domain Specific Language can be used in many ways. For
 example, we can build a simple infix pre--calculator for constants:
\input{schemeprog}

 And then, wherever we want to calculate a numerical constant in the 
compilation time, we may use the \scheme|exp1| macro:
\begin{schemedisplay}
 (exp1 5 + ((10 / 2)-(1 / 5)))
\end{schemedisplay}

 This language does not look like Scheme any more. And we can go even
further, implementing a Pascal (or Rlisp)--like language on top of Scheme,
using just the same \scheme|regexp| macro to describe both a lexer and a parser,
and then to compile the resulting code to the underlying Scheme.

\begin{verbatim}
(pasqualish
 "
 function fac(x)
 begin
   if (x > 0) then 
      x*fac(x - 1) 
   else 1;
 end
")
\end{verbatim}

 No more parenthesis that frighten non--Lisp programmers so much! Now
even Pascal programmers can use Scheme.

 The code samples above demonstrate some of the techniques available in
this approach. The complete implementation can be downloaded from \cite{dslengine}. It is possible
to produce not only languages with a computational model which is close
to the model of Scheme (eager dynamically typed functional languages
with imperative features), but any possible languages, providing 
small intermediate DSLs which simulate alternative computational
models. For those who need very lowlevel power it is possible to
produce an intermediate code in C language (for example, the
Bigloo Scheme \cite{bigloo} implementation allows to include C code 
when compiling
through C backend). For implementing complicated runtime models it is
easy to produce an intermediate Forth--like DSL on top of Scheme and
then use both Scheme and Forth metaprogramming powers.

\section{Alternatives}

 To make the picture complete, it is necessary to mention other possible choices for
 the Core Language. The popular programming language, C++, could
 become such a Core Language relatively easily. It has a
 Turing-complete macro system, unfortunately, featuring the language
 different from the host language (so only one stage preprocessing
 is possible). It lacks a good type system, but it could be simulated
 on top of the existing lowlevel features. There exist some
 implementations of the recursive descendant parsing combinators for
 C++ (e.g., Boost Spirit library \cite{boost}), implementation 
 of the functional programming (e.g., Boost Lambda \cite{boost}), 
 and even Lisp compilers on top  of the C++ template system. The
 runtime evaluation is available in different ways: using pluggable
 scripting languages other than C++, using the C++ interpreter \cite{cint}. 
 An interesting approach is described in \cite{tempo}.

 Another choice is Forth. It is a powerful metalanguage, but the
 core language remains too lowlevel and unsafe. Forth is often the
 only choice available for the embedded systems with limited
 resources.

 It is worth mentioning modern experimental extensions for 
 strictly typed functional languages: Template Haskell \cite{th} and
 MetaOCaml \cite{metaocaml}. Both of them conform well to all of 
 the Core Language
 requirements. Objective Caml also provides  one--stage
 metaprogramming using a sophisticated preprocessing engine CamlP4.
 And OCaml is quite good for implementing interpreters using the
 closure--based technique. Some examples can be found in \cite{vsl}.

 No doubt that Common Lisp would also be a very good platform since it shares
 almost all the features with Scheme with exception of 
 simplicity. The killing feature of Common Lisp is advanced
 runtime compilation in some of the major implementations 
(CMU CL \cite{cmucl} and its descendant SBCL \cite{sbcl} are good examples),
and the defmacro is guaranteed to be working in all the implementations 
available, which is a great advantage over Scheme.

 For relatively small projects Tcl \cite{tcl} would be a good choice.
Its computational model is based on rewrites (and primary data 
structures are just the strings of text), which renders an extremely powerful
metaprogramming tool. JavaScript language is also based on the rewrites
semantics, so it could be used for metaprogramming too.

\section{Conclusion}

 The idea of metaprogramming is not something esoteric. 
Metaprogramming is used widely by commercial 
programmers, they just do not realize it. 
The methodology proposed in this paper is an attempt of uncovering 
all the hidden power of the metaprogramming techniques available. 

The Scheme example presented above is part of the working project, which
already proved the supremacy of this approach. A subset of the Domain
Specific Languages hierarchy designed for the WWW data acquiring
project is shown on the Fig. 1.

The subject discussed requires future research and practical
approbation, whose final result may be
 a completely formalized, mathematically strict
methodology description and a Core Language which will best fit this
methodology.

\end{multicols}

\begin{figure}
\begin{center}
\scalebox{0.7}{\includegraphics{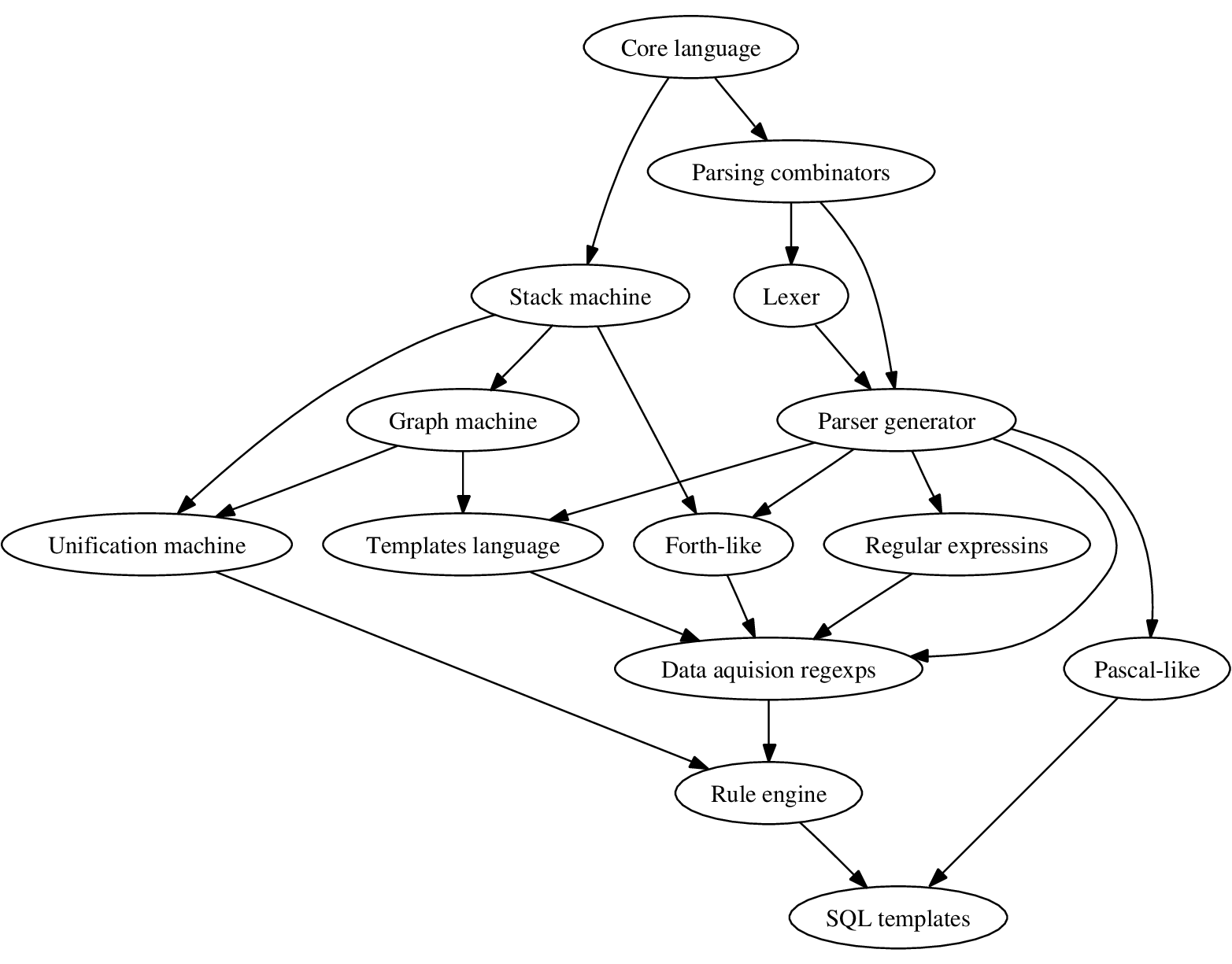}}
\end{center}
\caption{A sample DSLs hierarchy subset for the Web crawler project.}
\end{figure}

\end{document}

%% file: slinit.tex
\setspecialsymbol{lambda}{$\lambda$}
\setspecialsymbol{$0}{${}^\$0$}
\setspecialsymbol{$1}{${}^\$1$}
\setspecialsymbol{$2}{${}^\$2$}
\setspecialsymbol{$3}{${}^\$3$}
\setspecialsymbol{$4}{${}^\$4$}

\setspecialsymbol{p1}{$p_1$}
\setspecialsymbol{p2}{$p_2$}
\setspecialsymbol{r1}{$r_1$}
\setspecialsymbol{r2}{$r_2$}
\setspecialsymbol{px}{$p_x$}
\setspecialsymbol{pn}{$p_n$}
\setspecialsymbol{p+}{$p^+$}
\setspecialsymbol{p+0}{$p^{+0}$}
\setspecialsymbol{pOR}{$p^{OR}$}
\setspecialsymbol{pOR0}{$p^{OR^0}$}
\setspecialsymbol{po}{$\vec{p_o}$}
\setspecialsymbol{null?}{$\emptyset?$}

%% file: schemeprog.tex
\begin{schemedisplay}
(define-macro (exp1 . v)
 (defparsers
  (letrec
   ((epr
     (let
      ((body
        (regexp
         (num :-> $0) /
         (lst -> (aprs epr)))))
      (regexp
       ((body + (SCM psym +) + epr)
        :-> (list (+ $0 $2))) /
       ((body + (SCM psym -) + epr)
        :-> (list (- $0 $2))) /
       ((body + (SCM psym *) + epr)
        :-> (list (* $0 $2))) /
       ((body + (SCM psym /) + epr)
        :-> (list (/ $0 $2))) / body
        ))))
   (car (result (epr v))))))
\end{schemedisplay}